\documentclass[12pt]{iopart}

\usepackage{graphicx}
\usepackage{iopams}
\usepackage{revsymb}

\begin{document}

\title[ S Longhi, Optical realization of ...]{Optical realization of the two-site Bose-Hubbard model in waveguide lattices}

\author{S Longhi}

\address{Dipartimento di Fisica, Politecnico di Milano, Piazza L. da Vinci 32, I-20133 Milano,
Italy}
\ead{longhi@fisi.polimi.it}

\begin{abstract}
\noindent A classical realization of the two-site Bose-Hubbard
Hamiltonian, based on light transport in engineered optical
waveguide lattices, is theoretically proposed. The optical lattice
enables a direct visualization of the Bose-Hubbard dynamics in Fock
space.
\end{abstract}

\pacs{42.82.Et, 03.75.-b, 03.75.Lm}


\maketitle

The Bose-Hubbard Hamiltonian provides a paradigmatic theoretical
framework to investigate the physics of strongly interacting bosonic
systems \cite{BH}. In particular, the two-site Bose-Hubbard model
has received a great deal of attention in the last decade as a
simple model to investigate the dynamics of ultracold bosonic atoms
confined in a double-well potential
\cite{DW1,DW2,DW3,DW4,DW5,DW6,DW7}, the so-called bosonic junction
(for a recent review see \cite{reviewDW}). A remarkable property of
the bosonic junction is the existence of Josephson-like oscillations
of the atomic populations and of macroscopic self trapping due to
atom-atom interaction, which were originally predicted within a
semiclassical (mean-field) limit of the Bose-Hubbard model
\cite{DW1,DW2}. However, various quantum features of the
Bose-Hubbard dynamics,  such as quantum fluctuations, fragmentation
\cite{P1}, the cat-states formation in the supercritical attractive
case \cite{P2}, or the many-body coherent control of tunneling
\cite{CDT1,CDT2}, are not accessible within the mean-field limit.
This has lead to intensive studies on the relation between the full
quantum and mean-field dynamics (see, e.g., \cite{C0,C1,C2,C3} and
references therein). A comprehensive description of the entire
variety of phenomena rooted in the Bose-Hubbard Hamiltonian would
benefit from a direct access to quantum dynamical evolution in the
Fock space. However, in experiments with ultracold gases the
measurable quantities are usually atomic population imbalance and
relative phases \cite{DW2}, whereas full information on Fock state
occupation evolution is generally not accessible. \par In another
physical context, light transport in waveguide lattices has provided
a test bench to visualize in photonic systems the classical
analogues of a wide variety of coherent single-particle quantum
phenomena generally encountered in condensed matter or matter wave
systems \cite{Christodoulides03,Longhi09,Szameit10F}, such as Bloch
oscillations and Zener tunneling \cite{Christodoulides03,BO},
dynamic localization \cite{Longhi06,Szameit09}, and Anderson
localization \cite{and1,and2} to mention a few. As compared to
quantum systems, photonic lattices enable a direct and complete
visualization of the dynamics by mapping the temporal evolution of
the quantum system into spatial propagation of light waves in the
engineered lattice \cite{Longhi09}. Most of such previous
quantum-optical analogue studies, however, focused on
single-particle effects, whereas the possibility to visualize in a
classical optical setting the analogues of many-particle phenomena,
such as those rooted in the Bose-Hubbard model, remains unexplored.
Recently, photonic lattices with engineered coupling constants have
been proposed as classical analogs to quantum coherent and displaced
Fock states \cite{Szameit10}. \par In this Letter it is shown that
photonic lattices with suitably engineered coupling {\em and}
propagation constants provide a simple realization in the Fock space
of two-site Bose-Hubbard Hamiltonian.

\par
 The starting point of the analysis is provided by a standard
two-site Bose-Hubbard Hamiltonian that describes a system of $N$
interacting bosons occupying two weakly-coupled lowest states of a
symmetric double-well potential, which reads (see, for instance,
\cite{reviewDW})
\begin{equation}
\hat{H}=-\hbar J(\hat{a}^{\dag}_{1} \hat{a}_2+ \hat{a}^{\dag}_{2}
\hat{a}_1)+\frac{\hbar U}{2}(\hat{a}^{\dag \;2}_1
\hat{a}^{2}_1+\hat{a}^{\dag \;2}_2 \hat{a}^{2}_2)
\end{equation}
where $\hat{a}_{1,2}$ ($\hat{a}^{\dag}_{1,2}$) are the bosonic
particle annihilation (creation) operators for the modes in the two
wells, $J>0$ accounts for the coupling constant between the two
modes and $U$ is the strength of the on-site interaction ($U>0$ for
a repulsive interaction). If we expand the vector state of the
system $|\psi(t) \rangle$ on the basis of Fock states with constant
particle number $N$, i.e. after setting
\begin{equation}
| \psi(t) \rangle= \sum_{l=0}^N
\frac{c_l(t)}{\sqrt{l!(N-l)!}}\hat{a}^{\dag \; l}_{1} \hat{a}^{\dag
\; N-l}_{2} |0 \rangle
\end{equation}
the evolution equations for the amplitude probabilities $c_l(t)$ to
find $l$ particles in the left well and the other $(N-l)$ particles
in the right well, as obtained from the Schr\"{o}dinger equation $i
\hbar \partial_t | \psi(t) \rangle=\hat{H} | \psi(t) \rangle$, read
explicitly
\begin{equation}
i \frac{dc_l}{dt}=-(\kappa_lc_{l+1}+\kappa_{l-1}c_{l-1})+V_l c_l
\end{equation}
($l=0,1,2,...,N$), where we have set
\begin{equation}
\kappa_l=J \sqrt{(l+1)(N-l)} \; , \; V_l=\frac{U}{2}
\left[l^2+(N-l)^2-N \right].
\end{equation}
The normalization condition $\sum_{l=0}^N |c_l(t)|^2=1$ is assumed.
\par
 The evolution equations for the Fock state amplitudes $c_l$ can
be viewed as formally analogous to the coupled-mode equations
describing the transport of light waves in a tight-binding array
composed by $(N+1)$ waveguides with engineered propagation constant
shift $V_l$ and coupling rates $\kappa_l$ between adjacent
waveguides, in which the temporal evolution of the Fock-state
amplitudes of Bose-Hubbard Hamiltonian is mapped into the spatial
evolution of the modal field amplitudes of light waves in the
various waveguides along the axial direction $z$ (see, for instance,
\cite{Longhi09,Szameit10F}). The fractional light power distribution
$|c_l|^2$ in the various waveguides of the array at the propagation
distance $z$ thus reproduces the occupation probabilities of the
bosons between the two sites of the double-well. \par
In the optical
context, the tight-binding model (3) can be derived using a
variational procedure starting from the paraxial and scalar wave
equation for the electric field amplitude $\phi(x,z)$ describing the
propagation of monochromatic light waves at wavelength $\lambda$ in
an array of $(N+1)$ waveguides with refractive index profile $n(x)$
and substrate index $n_s$
\begin{equation}
i \lambdabar \partial_z \phi= -\frac{\lambdabar^2}{2n_s}\partial_x^2
\phi+[n_s-n(x)] \phi,
\end{equation}
\begin{figure}
\includegraphics[scale=0.6]{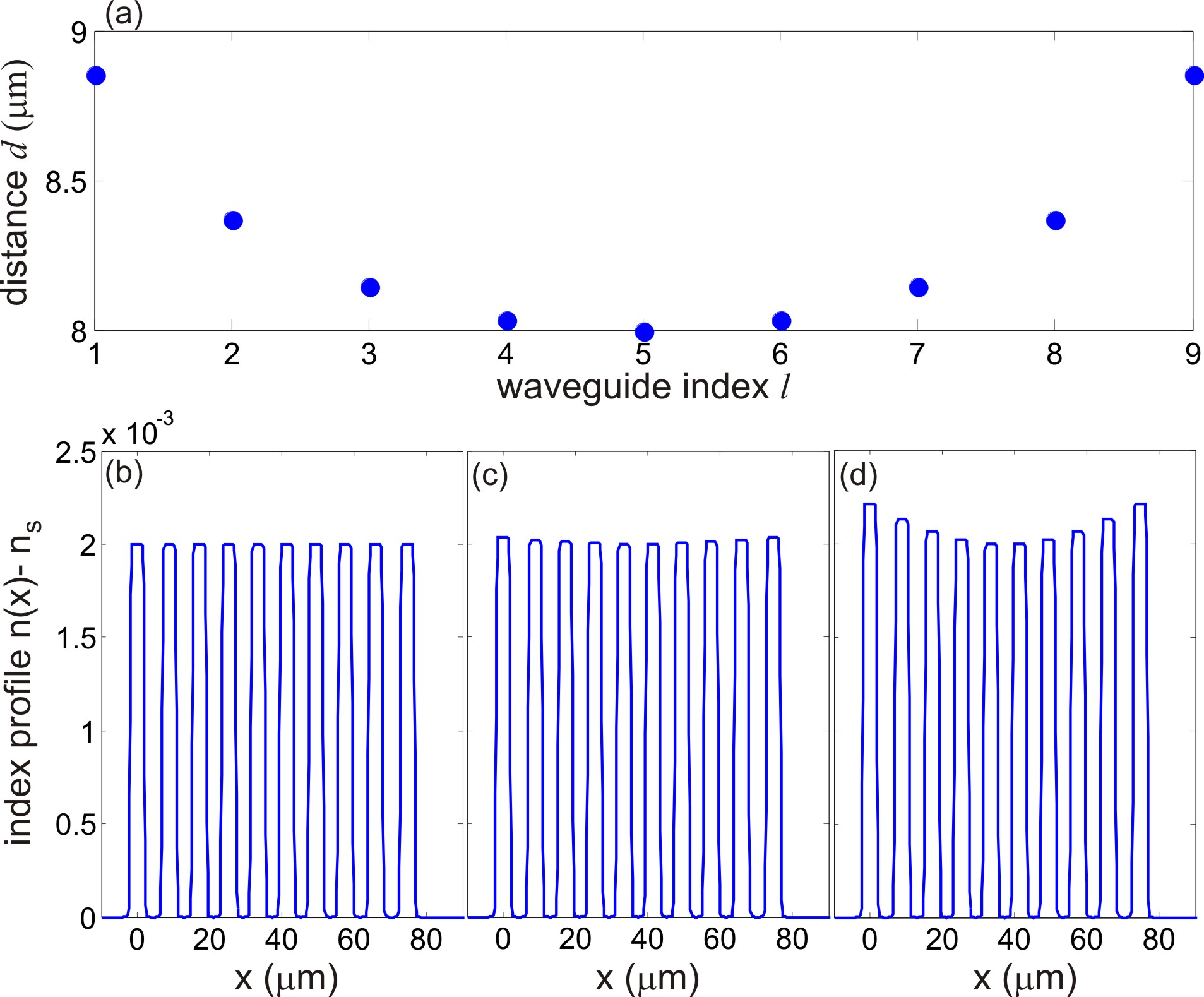}
\caption{ (color online) (a) Distribution of waveguide distance
($d_l$ is the distance between waveguides $(l-1)$ and $l$,
$l=1,2,...,9$) in a waveguide array composed by $N+1=10$ waveguides
that realizes the distribution of the coupling constants defined by
Eq.(4) for $J=0.0781 \; {\rm mm}^{-1}$. The panels (b-d) show the
refractive index profile $n(x)-n_s$ of the waveguide arrays that
realize the Bose-Hubbard Hamiltonian for increasing values of the
interaction strength $U$: in (b) $U=0$, in (c) $U=0.0174 \; {\rm
mm}^{-1}$, whereas in (d) $U=0.1043 \; {\rm mm}^{-1}$.}
\end{figure}
where $\lambdabar=\lambda/(2 \pi)$ is the reduced wavelength of
photons (for details see, for instance, \cite{Longhi06PLA}). In
particular, to independently engineer the coupling constants
$\kappa_l$ and propagation constant shifts $V_l$ of the waveguides,
one can assume a chain of waveguides with equal normalized
refractive index profile $g(x)$, but with different index contrasts
$\Delta n_l$ ($l=0,1,2,...,N$) and spacing $d_l=x_{l}-x_{l-1}$
($l=1,2,...,N$), i.e.
\begin{equation}
n(x)-n_s=-\sum_{l=0}^{N} \Delta n_l g(x-x_l).
\end{equation}
As discussed in the example below, to realize the Bose-Hubbard
lattice (3) the values of $d_l$ and $\Delta n_l$ slightly vary
around some mean values $d_r$ and $\Delta n$ that define a uniform
array. The waveguide separation $d_l$ mainly determines the value of
the coupling rate $\kappa_{l-1}$, with a characteristic exponential
dependence of $\kappa_{l-1}$ from $d_l$, whereas the index change
$\Delta n_l$ mainly defines the propagation constant mismatch $V_l$.
It is worth noticing that, in the absence of particle interaction,
i.e. for $U=0$, the lattice model (3) associated to the Bose-Hubbard
Hamiltonian was previously introduced in the photonic context to
realize exact spatial beam self-imaging in finite waveguide arrays
\cite{Gordon} and shown to belong to a rather general class of
exactly-solvable self-imaging tight-binding lattices with
equally-spaced energy levels \cite{Longhi10}. A nonvanishing value
of the interaction strength $U$ breaks the periodic self-imaging
property of the array and enables to visualize with light waves the
rich dynamical features of the two-site Bose-Hubbard Hamiltonian
directly in the Fock space (see, for instance,
\cite{DW1,DW2,DW3,DW4,DW5,DW6,DW7,reviewDW,C0,C1,C2,C3} and
references therein). Here two important dynamical features will be
considered for the sake of example, namely (i) the transition from
Josephson-like oscillations to self-trapping as the interaction $U$
is increased \cite{DW1,DW2,reviewDW}, and (ii) the transition from
single-atom to correlated pair tunneling of two bosons in a double
well potential \cite{DW7,fermi1}.\\
\begin{figure}
\includegraphics[scale=0.6]{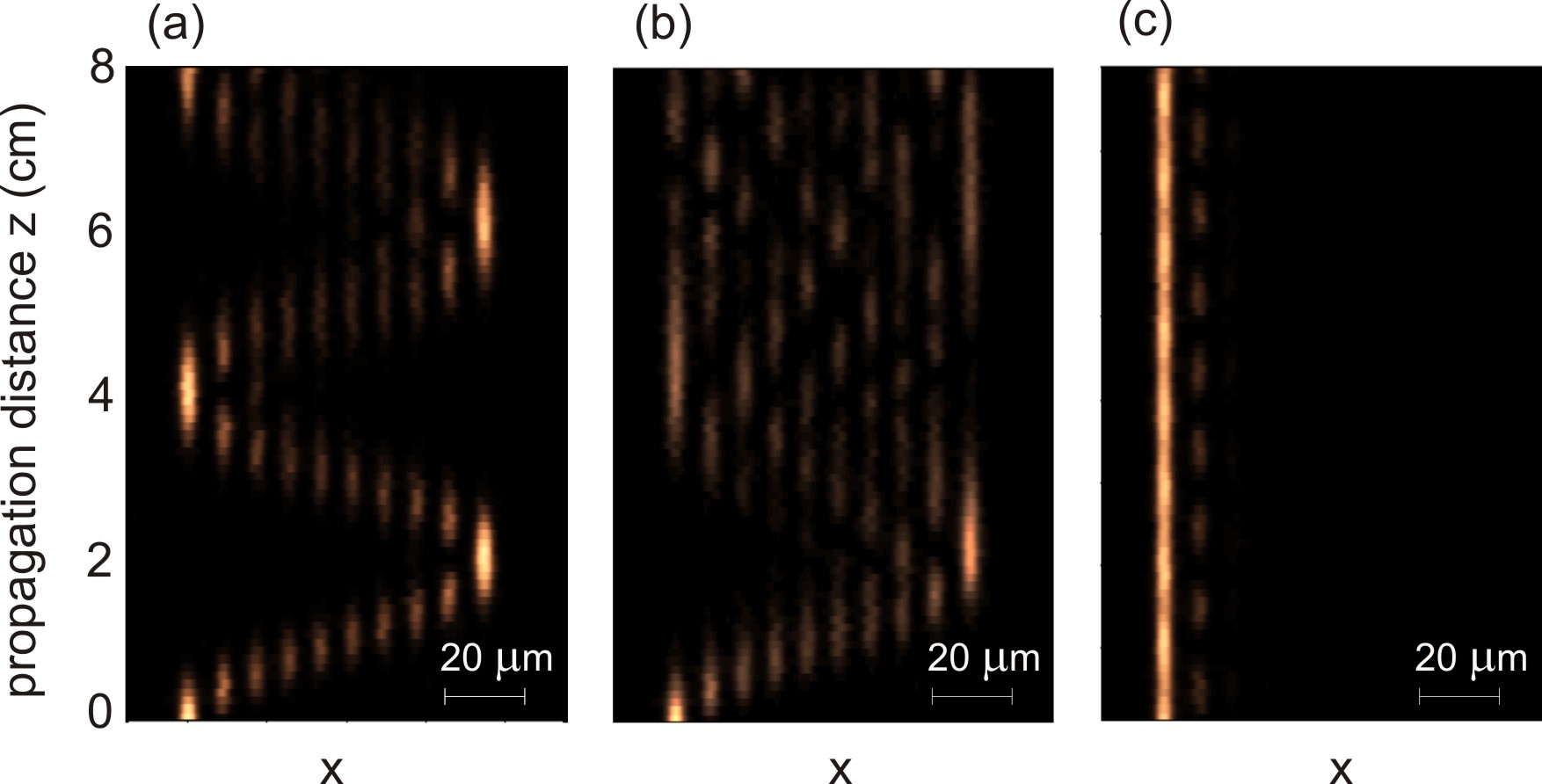}
\caption{ (color online) Light intensity evolution (snapshot of
$\phi(x,z)|^2$) in the three waveguide arrays of Figs.1(b-d)
corresponding to excitation of the left boundary waveguide, as
obtained by numerical simulations of the paraxial wave equation (5).
The images in (a), (b) and (c) correspond to the Bose-Hubbard
Hamiltonian for increasing values of the interaction strength $U$
(from left to right).}
\end{figure}
To visualize the first effect, we specifically design three arrays
comprising $N+1=10$ waveguides, which mimic the evolution of $N=9$
atoms in a double well, for three increasing values of the
interaction $U$. The arrays were designed to operate at
$\lambda=633$ nm in a transparent glass with a bulk refractive index
$n_s=1.45$ and with a normalized waveguide channel profile $g(x)=\{
{\rm erf}[(x+w)/D_x]-{\rm erf}[(x-w)/D_x] \}/[2 {\rm erf} (w/D_x)]$
with channel width $2w=4 \; \mu$m and diffusion length $D_x=0.3 \;
\mu {\rm m}^{-1}$. The distances between waveguides $d_l$ were
designed to realize the coupling rates $\kappa_l$ given Eq.(4) with
$J=0.0781 \; {\rm mm}^{-1}$. To determine the distribution of
distances $d_l$, a reference value $\Delta n=2 \times 10^{-3}$ of
refractive index contrast was assumed, and correspondingly the
coupling rate $\kappa$ between two adjacent waveguides versus
distance $d$ was computed, yielding to a good approximation the
exponential dependence $\kappa(d)=\kappa_0 \exp[-\gamma(d-d_r)]$ for
distances not too far from the reference distance $d_r=8 \; \mu$m,
where $\kappa_0 \simeq 0.3907 \; {\rm mm}^{-1}$ and $\gamma \simeq
0.6 \; \mu {\rm m}^{-1}$. The resulting distance distribution,
depicted in Fig.1(a), shows that the waveguide distances slightly
vary around the reference value $d_r$ indeed. By modulating the
index contrasts $\Delta n_l$ of the waveguides around the reference
value $\Delta n$, three different arrays were then designed to
realize the three different interaction regimes $U=0$, $U=0.0174 \;
{\rm mm}^{-1}$, and $U=0.1043 \; {\rm mm}^{-1}$, as shown in
Figs.1(b-d) \cite{note}. Such arrays could be fabricated in fused
silica by the recently developed femtosecond laser writing technique
\cite{Szameit10F}, in which the different refractive index contrasts
are obtained by varying the speed of the writing laser beam. Figures
2 show the evolution of light intensity $|\phi(x,z)|^2$ along the
three arrays, as obtained by numerical simulations of Eq.(5) using a
standard pseudospectral split-step method, when the left boundary
waveguide is excited at the input plane. Such an excitation
corresponds to the initial conditions $c_l(0)=\delta_{l,0}$, i.e. to
the entire $N$ bosons in the right well at the initial time. Note
that in case $U=0$ [Fig.2(a)] periodic self-imaging of the light
pattern is observed because of the equispacing of the energy levels
of the Bose-Hubbard Hamiltonian. Such periodic oscillations of the
light intensity pattern, previously predicted in Ref.\cite{Gordon}
and referred to as harmonic oscillations, are thus the optical
analogue of the bosonic Josephson oscillations in the
non-interacting regime \cite{DW1}. A quantitative measure of the
Josephson oscillations, generally used in the atom optics context,
is provided by the normalized population imbalance $P(z)$, which is
defined as the normalized difference between the average numbers of
atoms in the two wells, i.e. $P(z)=\sum_{l=0}^{N} [(N-2l)/N]
|c_l|^2$. Note that in an optical experiment $P(z)$ can be retrieved
 from a measurement of the beam center of mass position $\langle
 x(z) \rangle=\int dx x |\phi(x,z)|^2 / \int dx  |\phi(x,z)|^2$ using the simple relation $P(z)
\simeq 1-2\langle x(z) \rangle /(N d_r)$. For the $U=0$ case, the
evolution of $P(z)$ along the array is depicted in Fig.3(a) and
compared to the exact curve obtained from the tight-binding lattice
model (3). Note that $P(z)$ oscillates between -1 and 1 with spatial
period $z_R= \pi/J \simeq 4$ cm, which is precisely the period of
Josephson oscillations of the bosonic junction in the absence of
interaction. As the interaction strength is increased, the
self-imaging property of the array is clearly broken [see Figs. 2(b)
and (c)], with a clear transition to damped Josephson oscillations
[Fig.3(b)] and to the self-trapping regime [Fig.3(c)]. It should be
noted that the mean-field limit of the two-site Bose-Hubbard model
in the large $N$ limit, which is described by two nonlinear coupled
mode equations \cite{DW1,reviewDW}, could be realized in an optical
setting by a simple nonlinear optical directional coupler
\cite{Jensen}, and self-trapping phenomena in such nonlinear
couplers have been previously investigated. However, our waveguide
lattice system provides an exact realization of the Bose-Hubbard
model even for a low number of particles (as in the example previously discussed)
where the mean field approximation fails.\\
\begin{figure}
\includegraphics[scale=0.6]{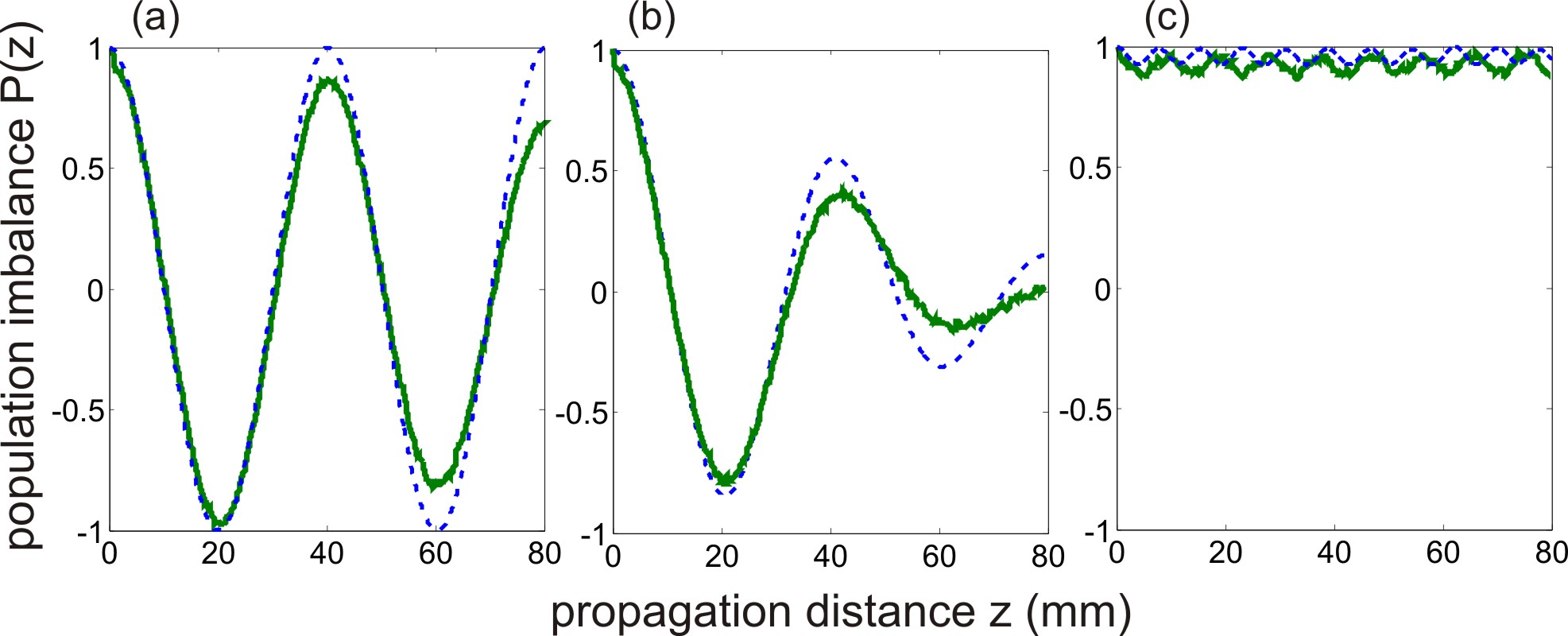}
\caption{ (color online) Behavior of the numerically-computed
population imbalance $P(z)$ for the three waveguide arrays of Fig.2
(solid curves), and corresponding behavior predicted by the
tight-binding Bose-Hubbard Hamiltonian (3). From (a) to (c), the
interaction strength takes the values $U=0$, $U=0.0174 \; {\rm
mm}^{-1}$, and $U=0.1043 \; {\rm mm}^{-1}$.}
\end{figure}

As a second example, we discuss the phenomenon of correlated pair
tunneling of two bosons in a double well potential induced by
interaction, which was investigated theoretically and experimentally
in previous works (see, for instance, \cite{DW7,fermi1}). Even if
the two-mode Bose-Hubbard model is capable of describing the
dynamics of such a system solely for weak interactions and far from
the fragmentation regime \cite{fermi1}, it well explains the
transition from Rabi oscillations of single particles to correlated
pair tunneling in the presence of a weak interaction. Such a kind of
correlated dynamics of a pair of interacting atoms was
experimentally observed  in Ref.\cite{DW7} by recording both the
atom position and phase coherence over time, and was explained on
the basis of the simplified Bose-Hubbard Hamiltonian (2). In the
$N=2$ particle case, the coupled-mode equations (3) reduce to the
following ones
\begin{eqnarray}
i \frac{dc_0}{dt} & = & -\sqrt{2} J  c_1+U c_0 \nonumber \\
i \frac{dc_1}{dt} & = & -\sqrt{2} J  (c_0+c_2) \\
i \frac{dc_2}{dt} & = & -\sqrt{2} J  c_1+U c_2. \nonumber
\end{eqnarray}
The tunneling dynamics can be analyzed by the introduction of the
percentage of bosons in the right well,
$p_R(t)=|c_0(t)|^2+(1/2)|c_1(t)|^2$, and the pair (or same-site)
boson probability $p_2(t)=|c_0(t)|^2+|c_2(t)|^2$ \cite{fermi1}, i.e.
the probability to find the two bosons in the same well (either the
left or the right well). The evolution of the particle occupation
probabilities $|c_l(t)|^2$ can be readily obtained by solving Eq.(7)
with the initial condition $c_0(0)=1$ and $c_1(0)=c_2(0)=0$. One
obtains
\begin{eqnarray}
|c_1(t)|^2 & = & \frac{2J^2}{M^2} \sin^2(Mt) \nonumber \\
|c_2(t)|^2 & = & \frac{1}{4} \left[
1+\cos^2(Mt)-2 \cos \left( \frac{Ut}{2} \right) \cos(Mt) \right. \nonumber \\
& -&  \left. \frac{U}{M} \sin \left( \frac{Ut}{2}\right)
\sin(Mt)+\frac{U^2}{4M^2} \sin^2(Mt) \right] \\
|c_0(t)|^2 & = & 1-|c_1(t)|^2-|c_2(t)|^2 \nonumber
\end{eqnarray}
where we have set $M=\sqrt{4J^2+U^2/4}$. Correspondingly, the
behavior of $p_R(t)$ and $p_2(t)$ reads
\begin{eqnarray}
p_R(t)& = & 1 -\frac{J^2}{M^2}\sin^2(Mt)- \frac{1}{4} \left[
1+\cos^2(Mt)-2 \cos \left( \frac{Ut}{2} \right) \cos(Mt) \right. \nonumber \\
& -&  \left. \frac{U}{M} \sin \left( \frac{Ut}{2}\right)
\sin(Mt)+\frac{U^2}{4M^2} \sin^2(Mt) \right] \\
p_2(t)& = & 1 -\frac{2J^2}{M^2}\sin^2(Mt)
\end{eqnarray}
\begin{figure}
\includegraphics[scale=0.6]{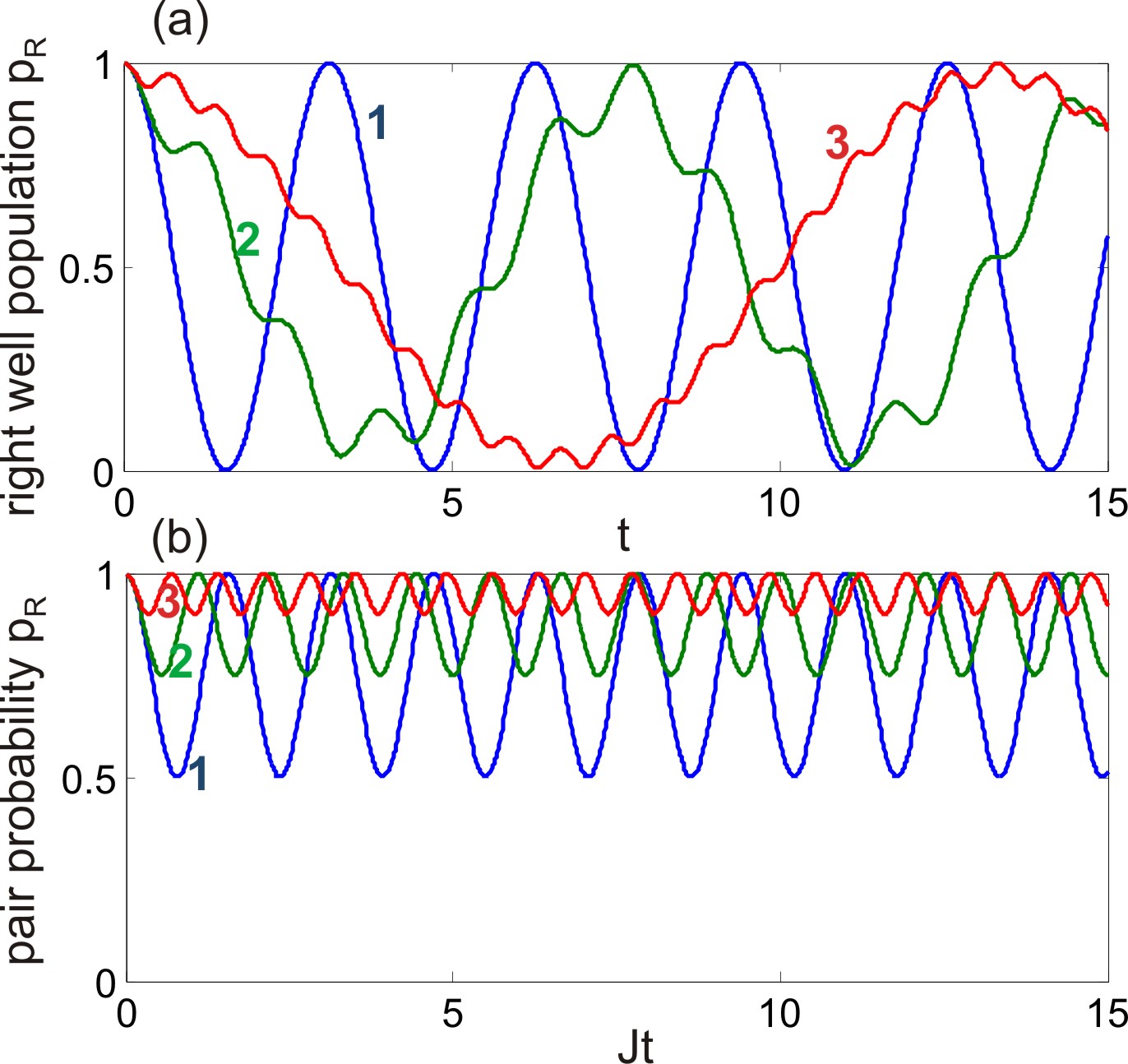}
\caption{ (color online) Behavior of (a) the percentage of bosons in
the right well $p_R(t)$, and (b) of the pair (same-site) boson
probability $p_2(t)$ in the Bose-Hubbard model with two atoms for
 for increasing values of the intercation strength $U$, normalized to the hopping rate $J$.
Curve 1: $U/J=0$; curve 2: $U/J=4$; curve 3: $U/J=8$.}
\end{figure}
A typical behavior of both $p_R(t)$ and $p_2(t)$ for increasing
values of the ratio $U/J$ is shown in Figs.4(a) and (b),
respectively. Note that, for the non-interacting case ($U=0$, curve
1 in Fig.4) the atoms simply Rabi oscillate back and forth between
both wells, and they tunnel independently. As a correlation is
introduced, the tunneling becomes a two-mode process and the
tunneling period increases [curves 2 and 3 in Fig.4(a)]. More
interestingly, from the behavior of $p_2(t)$ one can see that as the
interaction is increased both atoms remain essentially in the same
well in the course of tunneling, i.e they tunnel {\em as pairs} and
the amplitude $c_1(t)$ gets negligible \cite{fermi1}. In our optical
setting, such a dynamical behavior simply describes light tunneling
among three coupled waveguides, with the same coupling rate but with
the propagation constant of the central waveguide detuned from that
of the outer waveguides [see Eq.(7)], the detuning being
proportional to the interaction strength $U$ in the quantum
mechanical analogue. Such an optical structure was recently proposed
and realized for the observation of an optical analogue of
two-photon Rabi oscillations in Ref.\cite{Ornigotti08}. Indeed, the
large interaction regime $|U/J| \gg 1$ of the Bose-Hubbard model,
corresponding to $|c_1(t)| \ll 1$ and to pair tunneling (also
referred to as second-order tunneling \cite{DW7}), leads to Rabi
oscillations of light power between the outer waveguides of the
triplet system, with small excitation of the central waveguide.
\par In conclusion, a photonic realization of the two-site
Bose-Hubbard Hamiltonian in engineered waveguide lattices has been
proposed. Such an optical setting enables to visualize in the Fock
space the main dynamical aspects of interacting bosons. In
particular, waveguide lattices have been designed to visualize the
transition from Josephson-like oscillations to self-trapping, as
well as the transition from single-atom to correlated pair tunneling
in a simple two-boson system. It is envisaged that the idea proposed
in this work to use engineered waveguide lattices to simulate in a
purely classical setting the quantum physics of interacting
particles should motivate further theoretical and experimental
studies. For example, longitudinal modulation of the refractive
index in the lattice or the introduction of balanced loss and gain
in the waveguides might be used to mimic the recently proposed
kicked Bose-Hubbard \cite{CDT1} and non-Hermitian PT-symmetric
Bose-Hubbard models \cite{E2,E3}.

 This work was supported by the Italian MIUR (Grant
No. PRIN-20082YCAAK).


\begin{thebibliography}{99}
\bibitem{note}
A numerical computation of the propagation constant of the channel
waveguide indicates that a propagation constant shift $V_l$ is
approximately obtained by assuming a refractive index constrast
$\Delta n_l=\Delta n+ \beta V_l \lambdabar$, where $\Delta n= 2
\times 10^{-3}$ is the reference value of the index constrast and
$\beta$ is a numerical factor of order one ($\beta \simeq 1.2$ for
the chosen error function waveguide profile).
\end{thebibliography}
\end{document}